\documentclass[12 pt]{article}

\usepackage{amsmath,amssymb,amsfonts,latexsym}
\def\R{\mathbb{R}}

\begin{document}
%<<<<<<<<<<< ennumeration of eqns section wise>>>>>>>>>>>>>>>>>>>

\renewcommand\theequation{\arabic{section}.\arabic{equation}}
\catcode`@=11 \@addtoreset{equation}{section}
%<<<<<<<<<<<<<<<<<<<<<<<<<<<<<<<<<>>>>>>>>>>>>>>>>>>>>>>>>>>>>>>>>>
\newtheorem{axiom}{Definition}[section]
\newtheorem{axiom1}{Theorem}[section]
\newtheorem{axiom2}{Example}[section]
\newtheorem{lem}{Lemma}[section]
\newtheorem{prop}{Proposition}[section]
\newtheorem{cor}{Corollary}[section]
\newcommand{\be}{\begin{equation}}
\newcommand{\ee}{\end{equation}}

\newcommand{\equal}{\!\!\!&=&\!\!\!}
\newcommand{\lmat}{\left(\begin{array}{cccccc}}
\newcommand{\rmat}{\end{array}\right)}
\title{Chiellini integrability condition,  planar isochronous systems and Hamiltonian structures of
Li\'enard equation \\}

\author{A.Ghose Choudhury \footnote{Email: $aghosechoudhury@gmail.com$}\\
Department of Physics, Surendranath  College,\\ 24/2 Mahatma
Gandhi Road, Kolkata-700009, India.\\
\and
Partha Guha\footnote{E-mail: $partha@bose.res.in$}\\
S.N. Bose National Centre for Basic Sciences \\
JD Block, Sector III, Salt Lake \\ Kolkata - 700098,  India \\
}

 \maketitle
\begin{abstract}
Using a novel transformation involving the Jacobi Last Multiplier (JLM)
we derive an old integrability criterion due to Chiellini for the Li\'enard equation.
By combining the Chiellini condition for integrability and Jacobi's Last Multiplier  the Lagrangian and Hamiltonian of the Li\'enard equation 
is derived. We also show that the Kukles equation is
the only equation in the Li\'enard family which satisfies both the Chiellini integrability  and
the Sabatini criterion for isochronicity conditions. In addition we examine this result by mapping the
Li\'enard equation to a harmonic oscillator equation using tacitly Chiellini's condition. Finally we provide   a metriplectic and complex Hamiltonian formulation of the Li\'enard equation through the use of Chiellini condition for integrability.  \end{abstract}

\bigskip

{\bf Mathematics Classification (2010) :}

\smallskip


 34C14, 34C20
\section{Introduction}
A second-order nonlinear differential equation of the form $\ddot{x}+f(x)\dot{x}+g(x)=0$ where $f$ and $g$ are arbitrary real valued functions defined in an interval 
$I\subset \R$ is usually referred to as a Li\'{e}nard equation. There exists a vast literature on this equation owing to its widespread applications in several branches of the applied sciences. Equations of the Li\'{e}nard type are interesting because they display several novel features such as the existence of limit cycles, isochronicity, Hamiltonian structures etc.\\

A fundamental problem, common to any ordinary differential equation (ODE), is that of finding its solutions whenever they exist. As early as 1931 Cheillini \cite{Chiellini} 
in course of his investigations on the integrability of second-order  ODEs of the Li\'{e}nard type arrived at the following condition for their integrability, namely
\be\label{CC}\frac{d}{dx}\left(\frac{g}{f}\right)=sf\ee where $s$ is a constant.
This condition is today referred to as the Cheillini condition and as has been recently proved to be extremely useful in the construction of exact solutions after transforming the Li\'{e}nard equation to a first-order Abel equation of the first kind \cite{CG1,Harko1,Harko2,Kamke,MR1,MR2}.\\

On the other hand the Jacobi Last Multiplier (JLM), which was introduced in the context of a system of first-order ODEs more than 150 years ago by Carl Jacobi 
\cite{Jacobi. Jacobi1}, is also concerned with the possibility of reducing a system of first-order ODEs to a quadrature and is therefore essentially related to the issue of integrability. There is however another interesting application of the JLM specially for second-order ODEs. It may be shown that if the JLM is known for a second-order ODE (or equivalently for a first-order  planar differential system) then it is possible to deduce a Lagrangian for the system under consideration \cite{GCGK,NL2,NL3,NUT}. 
For the Li\'{e}nard equation it has been observed by the authors that the problem of finding a JLM and thereafter a Lagrangian for the equation is intimately related to the satisfaction of Cheillini's condition of integrability. This observation open up the door for investigation of the Hamiltonian  structures of second-order ODEs of 
the Li\'{e}nard type. A very promising way to algebrize the dynamics of a dissipative system is the metriplectic framework to which we shall be coming shortly. 
Recently much research \cite{CG,GC,RC}  has been devoted to the identification of isochronous dynamical systems whose motions are
completely periodic in their phase space. A center of a planar
dynamical system is said to be isochronous if all cycles near it
have same period.  For an equation of the Li\'{e}nard type the problem of determining the conditions on the functions $f$ and $g$ such that the center is isochronous was solved by Sabatini \cite{Sabatini,CD} who enunciated the requisite conditions to be satisfied by $f$ and $g$. Closely related to this problem is the question of finding a suitable transformation for mapping a planar dynamical system to that of a linear harmonic oscillator which is known to be isochronous. As the latter is integrable it is natural to enquire if in general the Cheillini condition, which also imposes conditions on $f$ and $g$, and the conditions necessary for isochronicity as deduced by Sabatini are compatible or not.\\

It is obvious that formally the Li\'{e}nard equation presents the structure of a dissipative system and as stated above one may investigate the use of the  metriplectic formalism to algebraize the dynamics of the equation.
 Metriplectic  systems were introduced by P.J. Morrison \cite{Morrison,Morrison1} and they combine both conservative and dissipative
systems. The dynamics of an isolated system with
dissipation is regarded as the sum of a Hamiltonian component, generated by $H$ via a
Poisson bracket algebra; plus dissipation terms, produced by a certain quantity, called entropy, $S$ via a
new symmetric bracket \cite{BKMR,Bloch,Grmela,Guha,Kaufman,Morrison1}. We also show how to express the Li\'enard equation in terms of the complex
Poisson Hamiltonian equation. In an interesting paper Rajeev \cite{Ra} has shown that
a large class of dissipative  systems can be
brought to a canonical form by introducing  complex co-ordinates in phase space and a
complex-valued Hamiltonian. In this paper we identify a class of dissipative systems
which yield complex Hamiltonization.

\bigskip

\smallskip

{\bf Result and organization}\, In this paper we derive an old result of an integrability criterion
of nonlinear differential equation, namely, the Chiellini integrability condition for the Li\'enard
equation, $\ddot{x} + f(x)\dot{x} + g(x) = 0$, using the Jacobi last multiplier. It is well known from the
work of many authors \cite{Harko1,Harko2,MR1,MR2,CG1,Mak,RMC} about the utility of the Chiellini condition for
solving nonlinear ODEs, but it is not known its role for the Hamiltonization of the Li\'enard
equation. In this paper we elucidate the construction of Lagrangian and Hamiltonian systems
of the Li\'enard equation using Jacobi's Last Multiplier. In particular, we obtain the bihamiltonian
structure of those Li\'enard equation which satisfies the Chiellini integrability condition
(\ref{CC}). By combining the Chiellini condition and Hamiltonization of the Li\'enard equation
we show that the Kukles equation is the only equation in the Li\'enard family which satisfies
both the Chiellini integrability and the Sabatini criterion for isochronicity conditions. Using
metriplectic dynamics we express the Li\'enard equation in complex Hamiltonian form with a
holomorphic Poisson structure.

\smallskip

This paper is {\bf organized} as follows. Using Jacobi's last multiplier technique and imposing
Cheillini integrability condition we  formulate the Lagrangian and
 Hamiltonian description of the Li\'enard equation in Section 2. In Section 3 we study isochronous Hamiltonian systems 
connected to Cheillini integrability condition. Section 4 is devoted to
metriplectic structure the  Li\'enard equation. We give a 
 complex Hamiltonian formulation of the  Li\'enard equation in Section 5.
We end this paper with a modest outlook.

\section{The Hamiltonian formulation of a Li\'{e}nard equation and Cheillini's integrability condition}

Consider a class of second order differential equations (ODE) in
which the damping term is proportional to the velocity $\dot{x}$, i.e.,
\be\label{k1}\ddot{x} + f(x)\dot{x}+g(x)=0 .\ee This is equivalent
to the standard system, \be \label{k2}\dot{x}=y,
\;\;\;\;\;\;\;\dot{y}= -f(x)y-g(x).\ee To deduce a Lagrangian for such a planar system we use
the  Jacobi Last Multiplier (JLM), $M$, whose
relationship with the Lagrangian, $L=L(t,x,\dot{x})$, for any
second-order equation of the form\be \label{k3} \ddot{x}=F(t,x,\dot{x})\ee is
\be \label{k4} M=\frac{\partial^2 L}{\partial \dot{x}^2}. \ee Note
that the JLM, $ M=M(t,x,\dot{x}) $ satisfies, by definition, the
following equation \be \label{k5} \frac{d}{dt}(\log M)+
\frac{\partial F}{\partial \dot{x}}=0 ,\ee which in the present
case is \be \label{k6} \frac{d}{dt}(\log M)-f(x)=0. \ee Assuming
that its formal solution is related to a new variable $u$ defined
\textit{via}
$$M(t,x)=\exp\left(\int f(x) dt\right):=u^{{1/\alpha}}$$ we find
that \be\label{3a} \dot{u}=\alpha u f(x).\ee Let us now set
$$\dot{x}=u+W(x),$$  where the particular form of $W(x)$ is to be determined. On  taking the time derivative of the last equation and using
(\ref{3a}) we have \be\label{3b}\ddot{x}=(\alpha
f(x)+W^\prime(x))\dot{x}-\alpha f(x) W.\ee Comparing (\ref{k1})
and (\ref{3b}) we see that
$$W^\prime(x)=-(\alpha+1)f(x)\hskip 20pt\mbox{and}\;\; \alpha
W(x)=\frac{g(x)}{f(x)}.$$ The consistency of these expressions
leads to the integrability condition:
\be\label{4a}\frac{d}{dx}\left(\frac{g}{f}\right)=-\alpha(\alpha+1)f(x),\ee
which serves to determine the parameter $\alpha$ for given $f$ and
$g$.
Comparison of (\ref{CC}) and (\ref{4a}) shows that the constant $s=-\alpha(1+\alpha)$.
Furthermore the system (\ref{k2}) is equivalent to the system
$$\dot{x}=u+\frac{1}{\alpha}\frac{g}{f}, \;\;\;\dot{u}=\alpha u f$$ subject of course to the condition (\ref{4a}).

 As already mentioned this integrability condition has
been employed for obtaining exact solutions of second-order
 differential equations that can be reduced to an Abel
 equation of the first kind.  A brief outline of the procedure is given below.

\subsection{Construction of an exact solution}
The third degree polynomial, first kind first-order
Abel differential equation is given by
\be\label{Abel}
\frac{dv}{dx} = f(x)v^2 + g(x)v^3,
\ee
where the coefficients $f(x)$ and $g(x)$  are real valued functions of
the variable $x$. The
Li\'enard system (\ref{k2}) can easily be reduced to a first-order equation
\be\label{Lie1}
y\frac{dy}{dx} = -f(x)y - g(x).
\ee
By the substitution $y = 1/v$, (\ref{Lie1}) can be transformed into a
first kind Abel differential equation of the form (\ref{Abel}).
Let $z = v g/f$,  then by using the Chiellini condition stated in (\ref{4a}) one obtains
\be\label{Abel1}
\frac{dz}{dx} = \frac{g}{f}\frac{dy}{dx} + \alpha f y = \frac{f^2}{g}(z^3 + z^2 - \alpha(\alpha+1) z).
\ee
This is separable and the solution is given by
\be
-\frac{1}{\alpha(\alpha+1)} \ln |\frac{g}{f} | = \int \frac{dz}{z(z^2 + z + \alpha)} + c,
\ee
where we have used  Chiellini's separability condition. This yields implicitly the solution of the
Li\'enard equation.

\smallskip

From the Chiellini condition (\ref{4a}) we have
\be\label{E1}
g = f(\beta - \alpha(\alpha+1)\int f(x)dx ) = K(x)f, \qquad \hbox{ where } \qquad K(x) = (\beta - \alpha(\alpha+1)\int f(x)dx ).
\ee
Thus it is clear that Chiellini condition implies that $g$ and $f$ are not independent and as such
the general Li\'enard equation and the corresponding Abel equation reduce to
\be\label{spLie}
\ddot{x} + f(x)\dot{x} + K(x)f(x) = 0, \qquad \frac{dv}{dx} = f(x)v^2(1 + Kv)
\ee
respectively. It is seen that  the restoring force $g(x)$ is proportional to the frictional force $f(x)$, where
the proportionality factor $K(x)$ is the primitive (potential) of $f(x)$. If we express
(\ref{spLie}) in terms of a potential function, then it becomes $\ddot{x} + f(x)\dot{x} + KK^{\prime} = 0$.
This yields
$$
\frac{1}{2}\frac{d}{dt} \big( \dot{x}^2 + K(x)^2) = - f(x)\dot{x}^2,
$$
which reflects for $f(x) \geq 0$, that the system is stable for all $x \neq 0$.
In the following we consider the Hamiltonian formulation of this particular form of the Li\'enard
equation.

\subsection{Chiellini integrability and Bi-Hamiltonian structure of poly-
nomial class of Li\'enard equation}

From the above discussion it is clear that the system (\ref{k2}) is  equivalent to the
following system \be\label{5a}\dot{u}=\alpha u f(x),\hskip 20pt
\dot{x}=u+\frac{1}{\alpha}\left(\frac{g}{f}\right),\ee  subject to Cheillini's condition for integrability (\ref{4a}). Assuming that this system admits a Hamiltonian structure we may recast (\ref{5a}) as
\be \label{k13}
\dot{x}=u+\frac{1}{\alpha}\frac{g}{f}=-J\frac{\partial{H}}{\partial
{u}},\;\;\;\;\;\;\;\;\dot{u}=\alpha u f(x)=J\frac{\partial
{H}}{\partial{x}},\ee where $H$ is the Hamiltonian of the system
and $J$ is symplectic, then up on equating the mixed derivative of
$H$ w.r.t. $x$ and $u$ we get the following linear partial
differential equation determining the symplectic $J$: \be
\label{k14} J_x(u+\frac{1}{\alpha}\frac{g}{f})+\alpha f(x)u
J_u=-Jf(x).\ee

%We assume that the function $f(x)$ is of the form
% \be\label{6a} f(x)=a\;x^\mu\ee then  condition
%(\ref{4a}) which then determines the function $g(x)$ leads to
%\be\label{6b}g(x)=x^\mu
%+a^2\frac{\alpha(1-\alpha)}{\mu+1}x^{2\mu+1},\ee where we have set
%the constant of integration to be $a^{-1}$, without loss of
%generality. Consequently the system (\ref{5a}) now appears as
%\be\label{7a}\dot{u}=\alpha a u x^\mu,\ee
%\be\label{7b}\dot{x}=(u+\frac{1}{\alpha a})
%+\frac{1-\alpha}{\mu+1} a x^{\mu+1}.\ee

%In general if we assume \be \label{k13}
%\dot{x}=u+\frac{1}{\alpha}\frac{g}{f}=-J\frac{\partial{H}}{\partial
%{u}},\;\;\;\;\;\;\;\;\dot{u}=\alpha u f(x)=J\frac{\partial
%{H}}{\partial{x}},\ee where $H$ is the Hamiltonian of the system
%and $J$ is symplectic, then up on equating the mixed derivative of
%$H$ w.r.t. $x$ and $u$ we get the following linear partial
%differential equation determining the symplectic $J$: \be
%\label{k14} J_x(u+\frac{1}{\alpha}\frac{g}{f})+\alpha f(x)u
%J_u=Jf(x).\ee

% For the chosen form of $f(x)$ and $g(x)$ given in
%(\ref{6b}) this becomes \be\label{8a} J_x\left[(u+\frac{1}{\alpha
%a }+\frac{1-\alpha}{1+\mu}a x^{\mu+1}\right]+(a\alpha u
%x^\mu)J_u=J a x^\mu.\ee
The  Lagrange system for
(\ref{k14}) is in general \be\label{k14a}
\frac{dx}{u+\alpha^{-1}g/f}=\frac{du}{\alpha u
f}=\frac{dJ}{-Jf}.\ee
%In the specific case under consideration
Its characteristics are easily found to be:
\be\label{k14b}c_1=J u^{1/\alpha}\ee
 \be\label{k14c}c_2=u^{(\alpha+1)/\alpha}\left[\frac{g}{f}+\frac{\alpha(\alpha+1)}{2\alpha+1}u\right],\ee
where $J$  is the component of a
symplectic matrix.
The general solution of (\ref{k14a}) is therefore of the form $c_1=F(c_2)$ where $F$ is an arbitrary function. Assuming $F(c_2)=c_2$ we have
\be\label{x1}J=u\left(\frac{g}{f}+\frac{\alpha(\alpha+1)}{2\alpha+1}u\right).\ee
It remains to calculate the Hamiltonian $H$ which using the last expression for $J$ in (\ref{k13}) is given by
\be\label{x2}H=\ln \left[|u|^{-1/\alpha}\left|\frac{g}{f}+\frac{\alpha(\alpha+1)}{2\alpha+1}u\right|^{-1/(\alpha+1)}\right].\ee
%Thus we claim:
%\begin{prop}
%Let the Li\'enard equation satisfy Chiellini's integrability condition
%then the planar system $\dot{u} = \alpha uf(x),
%\dot{x} = u + \frac{1}{\alpha}\frac{g}{f}$ admits a  bihamiltonian structure,
%with symplectic structures  given by
%\be
%J_1 = k_1u^{1/\alpha}, \qquad J_2 = k_2 u^{1/(1 - \alpha)},
%\ee
%where $k_1$ and $k_2$ are arbitrary constants.
%\end{prop}

\subsection{A Lagrangian and Hamiltonians of Li\'enard equation}

Now from (\ref{k4}) and (\ref{5a}) we have
$$\frac{\partial^2L}{\partial
\dot{x}^2}=\left(\dot{x}-\frac{1}{\alpha}\frac{g}{f}\right)^{1/\alpha},$$ so
that
$$L(x,\dot{x},t)=\frac{\left(\dot{x}-\frac{1}{\alpha}\frac{g}{f}\right)^{1/\alpha+2}}{(1/\alpha+1)(1/\alpha+2)}
+h_1(x,t)\dot{x}+h_2(x,t).$$ Here $h_1(x,t)$ and $h_2(x,t)$ are
arbitrary functions of integration. Inserting this Lagrangian into
the Euler-Lagrange equation and using (\ref{k1}) we find that
$$h_{1t}-h_{2x}=0.$$ Therefore choosing $h_1(x,t)=G_x$ and
$h_2(x,t)=G_t$ it follows that \be
L=\frac{\left(\dot{x}-\frac{1}{\alpha}\frac{g}{f}\right)^{1/\alpha+2}}{(1/\alpha+1)(1/\alpha+2)}
+\frac{dG}{dt}.\ee We can drop the total derivative term without
loss of generality. The conjugate momentum is then defined in the
usual manner by
$$p=\frac{\partial L}{\partial
\dot{x}}=\frac{\left(\dot{x}-\frac{1}{\alpha}\frac{g}{f}\right)^{1/\alpha+1}}{(1/\alpha+1)},$$
and therefore
$$\dot{x}=\frac{1}{\alpha}\frac{g}{f}+\left((1/\alpha+1)p\right)^{1/(1/\alpha+1)}.$$
Using the standard Legendre
transformation the Hamiltonian is found to be \be\label{Hamlie} H=p\dot{x}-L=\frac{1}{\alpha}
p\frac{g}{f}
+\frac{\alpha}{2\alpha+1}\left(\frac{\alpha+1}{\alpha}p\right)^{\frac{2\alpha+1}{\alpha+1}}.\ee The corresponding canonical
equations are: \be\dot{x}=\frac{\partial H}{\partial
p}=\frac{1}{\alpha}
\frac{g}{f}+\left(\frac{\alpha+1}{\alpha}p\right)^{\frac{\alpha}{\alpha+1}},\ee
\be\dot{p}=-\frac{\partial H}{\partial x}=(\alpha+1)f p.\ee
 In terms of the   scaled variable $\tilde{p}:=(\alpha+1)/\alpha p$  the Hamiltonian (\ref{Hamlie}) has the following
 appearance,
 \be\label{Hamlie2}H(x,\tilde{p}; \alpha)=\frac{1}{\alpha+1}\tilde{p}\frac{g}{f}+
 \frac{\alpha}{2\alpha+1}\tilde{p}^{\frac{2\alpha+1}{\alpha+1}}.\ee 
Note that $\alpha$ is a parameter and $H$ changes if $\alpha$ changes.
The canonical
 Poisson bracket accordingly becomes $\{x,\tilde{p}\}=(\alpha+1)/\alpha$.
 From (\ref{E1})  the function $g=K(x)f$ and hence
  the
 Hamiltonian assumes the form
 $$H(x,\tilde{p};\alpha)=\frac{\tilde{p}}{\alpha+1}K(x)+\frac{\alpha}{2\alpha+1}\tilde{p}^{\frac{2\alpha+1}{\alpha+1}}.$$

\section{Isochronicity and the Chiellini integrability condition}

In this section we first identify the isochronous cases resulting
from the Li\'enard equation and examine the corresponding
Hamiltonian structures.
It is shown by  Sabatini that if the functions $f$ and $g$ be analytic, $g$
odd, $f(0)=g(0)=0$ and $g^\prime(0)>0$ then the origin is an
isochronous center if and only if $f$ is odd and
\be\label{tau}\tau(x):=\left(\int_0^x s f(s) ds\right)^2-x^3
(g(x)-g^\prime(0)x)\equiv 0.\ee

With out loss of generality we assume $g^\prime(0) =1$.
Thus in case of isochronicity the function $g(x)$ has a specific
form depending on $f(x)$ as given by (\ref{tau}). If in (\ref{tau}) we substitute the expression for $g(x)$ resulting from the Cheillini condition, i.e., $g(x)=K(x) f(x)$ with $K(x)=\beta -\alpha(1+\alpha)\int_0^x f(s) ds$, we obtain
\be\label{4.2} x+\frac{1}{x^3}\left(\int_0^x s f(s) ds\right)^2=g(x)=\left(\beta-\alpha(\alpha+1)\int_0^x f(s) ds\right) f(x).\ee
Assuming $f(x)=kx^\mu$ its substitution into (\ref{4.2}) gives

$$x+\frac{k^2}{(\mu+2)^2}x^{2\mu+1}=-\alpha(\alpha+1)\frac{k^2}{\mu+1}x^{2\mu+1}+\beta kx^\mu.$$
Choosing $\beta=1/k$, we are therefore led to conclude that $\mu=1$.
Consequently it follows that the parameter $\alpha$ has to satisfy
the following equation, $9\alpha^2+9\alpha+2=0$, and its solutions
are $\alpha=-1/3$ and $\alpha=-2/3$ respectively.\\
Hence the   Li\'{e}nard equation admits isochronous motion only
when $f(x)=kx$ and $g(x)=x+k^2x^3/9$. Furthermore with the above
values of the parameter $\alpha$ we see that the Hamiltonians,
from (\ref{Hamlie2}), are as follows:

$$H_1(x,\tilde{p},\alpha=-1/3)= \frac{3}{2}\frac{g}{f}\tilde{p}-\tilde{p}^{1/2}, \qquad \hbox{ where } \qquad
{\tilde p} = - 2p,$$
$$H_2(x,\tilde{p},\alpha=-2/3)=\frac{2}{\tilde{p}} + 3\frac{g}{f}\tilde{p},  \qquad \hbox{ where } \qquad
{\tilde p} = - \frac{1}{2}p,$$
where $g/f=1/k+kx^2/9$. Both the Hamiltonians give rise to the
same equation, namely
\be\label{iso1}\ddot{x}+kx\dot{x}+\left(x+\frac{k^2}{9}x^3\right)=0,\ee
which reduces to the standard Kukles system for the choice $k=3$.
The Poisson brackets corresponding Hamiltonian structures $H_1$ and $H_2$ are given by
$\{x,\tilde{p}\}=-2$ and $\{x,\tilde{p}\}=-\frac{1}{2}$ respectively.
Thus it is seen that the isochronous equation (\ref{iso1}) admits
a bi-Hamiltonian structure.

  Thus we have shown that for the Li\'{e}nard type equations, when
  the coefficient of the damping term is  linear in $x$, then
  there is only one equation having an isochronous center, namely
  (\ref{iso1}).

\subsection{Mapping of a Kukles system to the harmonic/isotonic oscillator equation and isochronicity}

In this section
we examine the relationship between (\ref{iso1}) and the equation of a linear harmonic and the isotonic oscillator.
These are the only two systems with a rational potential which display the property of isochronoicity.
 The system (\ref{iso1}) is equivalent to
 \be\label{Y1} \dot{u}=\alpha u k x,\;\;\;\;\dot{x}=u+\frac{1}{\alpha k} +\frac{k}{9\alpha} x^2,\ee
 with $9\alpha^2+9\alpha +2=0$. The associated symplectic structure may be taken as
 $J=cu^{-1/\alpha}$ for the allowed values of $\alpha=-1/3$ and $-2/3$ .
 The Hamiltonian for (\ref{Y1})  with this symplectic form turns out to be
 \be\label{Y2}H=\frac{\alpha k}{2c} x^2 u^{(\alpha+1)/\alpha}+V(u) +s,\ee
 where $s$ is a constant and the potential
 \be\label{Y3} V(u, \alpha)=-\frac{1}{c k (\alpha+1)}u^{(\alpha+1)/\alpha}-\frac{\alpha}{c(2\alpha+1)}u^{(2\alpha+1)/\alpha}.\ee
 \textbf{Case 1:} When $\alpha=-1/3$ the expression for the potential simplifies to\\
 $$V(u, \alpha=-1/3)=-\frac{3}{2ck}\left(\frac{1}{u}-\frac{k}{3}\right)^2+\frac{k}{6c}.$$
  By defining a new set of variables
 $\xi:=\frac{x}{u}$ and $\eta:=\frac{1}{u}-\frac{k}{3}$
 the Hamiltonian may be expressed as
 $$H_{h.o}=-\frac{k}{6c}\xi^2-\frac{3}{2ck}\eta^2$$
 after setting $s=-k/6c$.  
 The overall negative sign can be fixed by an appropriate choice of the values of $k$ and $c$. For instance if $k=3$ and $c=-1$ then the above Hamiltonian reduces to the standard form
 \be H_{h.o}=\frac{1}{2}\xi^2+\frac{1}{2}\eta^2, \;\;\;\{ \xi, \eta\}=-1.\ee
\textbf{Case 2:} For $\alpha=-2/3$ the choice
 $\xi:=\frac{x}{u^{1/4}},\;\;\;\eta:=\frac{2}{u^{1/4}}$
 causes the Hamiltonian to become
$$H_{iso}=-\frac{k}{3c}\xi^2-\frac{2}{c}\frac{4}{\eta^2}-\frac{3}{ck}\frac{\eta^2}{4},$$
 which upon choosing $k=3, c=-2$ reduces to the form
 \be H_{iso}=\frac{1}{2}\xi^2 +\frac{1}{8}\eta^2+\frac{4}{\eta^2}, \;\;\;\{\xi, \eta\}=-1, \ee
and  corresponds to that of the isotonic oscillator. The latter is also known as the singular harmonic oscillator as the singular term has the effect of  centrifugal type of potential. The classical motion is confined either to the region $\eta>0$ or $\eta<0$ and the corresponding quantum system is Schr\"{o}dinger solvable with an equispaced spectrum.

 %Setting $k=-3$ it is possible to rewrite (\ref{iso1}) as the following system
%\be\label{L1} \dot{u}=x u,\;\;\;\dot{x}=u+x^2+1\ee
%To recast (\ref{L1})  in the form of the equations of a linear harmonic
% oscillator we define a new set of variables $\xi$ and $\eta$
% $$\xi=\frac{x}{u}, \eta=\frac{1}{u}+1$$
%A simple calculation shows that
%$$\dot{\xi}=\eta, \;\;\;\dot{\eta}=-\xi$$ which are identical to the equations for linear harmonic oscillator. This $further illustrates the reason for the isochronous behaviour of the nonlinear equation (\ref{iso1}).

\subsubsection{Solution via Chiellini integrability condition}

We complete this article by obtaining the solution of the Kukles equation using
Chiellini integrability condition.
Once again consider  the L\'ienard equation in the form
\be\label{Z1} \ddot{x} + \lambda x \dot{x} + \beta x^3 + \gamma x = 0 \ee
Here $f(x)=\lambda x$ and $g(x)=\gamma x+\beta x^3$ and
the  Chiellini condition is given by
$$
\frac{d}{dx}\big(\frac{g(x)}{f(x)}\big) = \frac{2\beta}{\lambda^2}f(x).
$$
Set $\dot{x} = c_k \frac{g(x)}{f(x)}$, where $c_k$ is a constant to be determined, and differentiate this once more to obtain
$$
\ddot{x} = c_{k}^{2}\frac{2\beta}{\lambda^2} g(x),
$$
where we have used the Chiellini condition.
Substituting the first and second-order derivatives of $x$ into (\ref{Z1}) we have
$$
c_{k}^{2}\frac{2\beta}{\lambda^2}g(x) + f(x)c_k\frac{g(x)}{f(x)} + g(x) = 0,
$$
which immediately determines the constant $c_k$ as
\be
c_k = \frac{-\lambda^2 \pm \sqrt{\lambda^2 - 8\beta^2}}{4\beta}.
\ee
As $ g(x) = \beta x^3 + \gamma x$ and $f(x) = \lambda x$, it follows from $\dot{x} = c_k \frac{g(x)}{f(x)}$
that
\be
\frac{1}{\sqrt{\frac{\gamma}{\beta}}}\tan^{-1} \frac{x}{\sqrt{\frac{\gamma}{\beta}}}
= \frac{\beta}{\lambda}c_k (t - t_0).
\ee

\section{Chiellini integrability condition and metriplectic structure}

In this section we show that using Chiellini  integrability condition the  Li\'{e}nard equation
can be reformulated in terms of the complex Hamiltonian theory. To this end we define
\be V_x = \frac{g}{f}, \qquad V_{xx} \equiv
\frac{d}{dx}(\frac{g}{f}) = \mu f. \ee

Therefore in terms of the new function $V$ a Li\'{e}nard type ODE
has the form therefore be written as \be \ddot{x} +
\frac{1}{\mu}V_{xx}\dot{x} + (\frac{1}{2\mu}V_{x}^{2})_x = 0. \ee

As a  first-order system of  ODEs it may be recast as

\begin{eqnarray}\label{newLienard}
\dot{x} \equal y \;, \nonumber \\
\dot{y} \equal -\frac{1}{\mu}V_{xx}y -(\frac{1}{2\mu}V_{x}^{2})_x
\;.
\end{eqnarray}

\begin{lem}
The Lienard equation transforms under \be y = p - V_x. \ee to new
set of first order ODE
\begin{eqnarray}
\dot{x} \equal p - V_x \;, \nonumber \\
\dot{p} \equal -(\frac{1}{\mu} - 1)V_{xx}p
-(\frac{1}{2\mu}V_{x}^{2})_x  \;.
\end{eqnarray}
\end{lem}

{\bf Proof}: It is clear that
$$
\dot{p} = V_{xx}(p-V_x) - \frac{1}{\mu}V_{xx}(p-V_x) -
\frac{1}{2\mu}(V_{x}^{2})_x
$$
$$
=  -(\frac{1}{\mu} - 1)V_{xx}p -(\frac{1}{2\mu}V_{x}^{2})_x.\qquad
\Box$$

Our aim is next to rewrite the system (\ref{newLienard}) in a metriplectic and complex form.
Let $S$ be a real valued function on a $m$-dimensional manifold $M$. If $M$ is
compact smooth Riemannian manifold, the gradient vector field associated with the metric
$g = \sum g_{ij}dx^i \otimes dx^j$ is given by
$$
\hbox{ grad}(S) = G\big(\frac{\partial S}{\partial x_1}, \cdots , \frac{\partial S}{\partial x_m \big)},
$$
where $G = (g_{ij})$ and $(x_1, \cdots , x_m)$ is a local coordinate.

\bigskip

P.J. Morrison \cite{Morrison} introduced a natural geometrical formulation of dynamical systems
that exhibit both conservative and nonconservative characteristics.
A metriplectic system \cite{BKMR,Bloch,Grmela,Guha,Kaufman,Morrison} consists of a smooth manifold $M$ , two
smooth vector bundle maps $J^{\sharp}, G^{\sharp} : T^{\ast}M \to TM$ covering the identity,
and two functions $H, S \in C^{\infty}(M)$, the Hamiltonian or total energy
and the entropy of the system, such that it yields Poisson bracket and
positive semidefinite symmetric
bracket
$$
J(df,dh) = \{f,h\}, \qquad G(df,dh) = (f,g),
$$
respectively. Moreover, the additional requirements that $H$
remains a conserved quantity and $S$ continues to be dissipated. These requirements can be met if the
following conditions on $H$ and $S$ are satisfied
$\{S,F\} = 0$ and $(H,F) = 0$ for all $F \in C^{\infty}(M)$,
i.e, $JdS = GdH = 0$. It shows that $S$ is a Casimir function for the Poisson tensor
$J$ and $dH$ is a null vector for the symmetric tensor $G$.

In this paper we work with a slightly weaker condition of metriplectic
condition, i.e. $JdS + GdH = 0$.

\begin{prop}
The Li\'enard equation of motion take the following form
\be
\dot{X} = J\nabla H_1 - G \nabla S,
\ee
where $ X = \left( \begin{array}{c}
x \\
p \\
\end{array} \right)$.
Here $J$ is the standard symplectic matrix and the second term represents gradient flow, where  $G$ is
defined by
$$
G = \left(
\begin{array}{cc}
\frac{1}{\alpha} & 0 \\
0 & \alpha \\
\end{array}
\right),
$$
where $\alpha$ is a parameter. The $H_1$ and $S$ are given by
\be\label{h1} H_1 = \frac{1}{2}p^2 + \frac{1}{2\mu} V_{x}^{2} +
\Big((\frac{1}{\mu} - 1)V_x - \alpha x \Big)p, \ee \be
\label{h2}S = \frac{1}{2}p^2 + \alpha \big(\frac{1}{\mu}V -
\frac{\alpha}{2}x^2\big). \ee
\end{prop}

{\bf Proof}: It is easy to see that \be H_{1x} =
\Big[\frac{1}{\mu}V_xV_{xx} + \big((\frac{1}{\mu} - 1)V_{xx} -
\alpha  \big)p \Big], \ee \be H_{1p} = p +
\big((\frac{1}{\mu} - 1)V_x - \alpha x \big), \ee \be
S_{x} = \alpha [\frac{1}{\mu}V_x - \alpha x], \ee and
$S_{p} = p$. Using all these expressions we obtain our result.
$\Box$

\bigskip

\begin{cor}
If $\mu = 2$ and $ p = V_x$ then the Li\'enard equation satisfies weaker metriplectic condition, i.e.,
$JdS + GdH = 0$.
\end{cor}

\section{Complex Hamiltonian formulation and Li\'{e}nard equation}

Suppose $S$ be the symplectic foliation of $M$. We denote by $N$ the distribution
defined as the $g$-orthogonal complement to $S$. Thus at every point $m$ a decomposition
into direct sum of sub-bundles, i.e. $T_mM = T_mS \oplus N_x$. If the Poisson bivector $\Pi$ is parallel
with respect to the Levi-Civita connection $\nabla$, i.e. $\nabla \Pi = 0$. There is a classical result
of Lichnerowicz \big(\cite{Lich},(\cite{Vaisman}, Remark 3.11)\big) that the distribution $N$ is integrable.
Hence together with the symplectic structure and
the restriction of the metric $g$ to the symplectic leaves defines a K\"ahler structure.

\bigskip

It is also possible to express the Hamiltonian of an equation of the  Li\'{e}nard type
  within the framework of the complex Hamiltonian theory. In this section we adopt more straight forward approach,
and once again we demostrate the role of Chiellini integrability condition.

\begin{prop}
The equations of motion take the complex form, given by
\be
\frac{d}{dt}\left( \begin{array}{c}
p \\
x \\
\end{array}
\right) = \left( \begin{array}{c}
\{H_1, {p} \} \\
\{ H_1, x \} \\
\end{array}
\right) + J\left( \begin{array}{c}
\{H_2, { p} \} \\
\{ H_2, x \} \\
\end{array}
\right),
\ee
where $J$ is an almost complex structure
defined by
$$
J = \left(
\begin{array}{cc}
0 & -\alpha \\
\frac{1}{\alpha} & 0 \\
\end{array}
\right),
$$
where $\alpha$ is a parameter. The Hamiltonians are given by
\be\label{h1} H_1 = \frac{1}{2}p^2 + \frac{1}{2\mu} V_{x}^{2} +
\Big((\frac{1}{\mu} - 1)V_x - \alpha x \Big)p, \ee \be
\label{h2}H_2 = \frac{1}{2}p^2 + \alpha \big(\frac{1}{\mu}V -
\frac{\alpha}{2}x^2\big). \ee
\end{prop}

A complex structure allows one to endow a real vector space
$\cal{V}$ with the structure of a complex vector space. In other
words, given any real vector space $\cal{V}$ we may define its
complexification by $\cal{V}^{\Bbb C} = \cal{V} \otimes_{\Bbb
R}{\Bbb C}$ and $J$ is guaranteed to have eigenvalues which
satisfy $a^2 = -1$, namely $a = \pm i$. Thus we may write
$$
\cal{V}^{\Bbb C} = \cal{V}^+ \oplus \cal{V}^-
$$
where $\cal{V}^+$ and $\cal{V}^-$ are the eigenspaces of $+i$ and
$-i$ respectively. Given such a matrix $J$ we can define the
equation of motion in terms of the complex coordinates \be \dot{z}
= \{H_{\Bbb C}, z \}, \ee
 generated by
the complex Hamiltonian function $ H_{\Bbb C} = H_1 + iH_2$, where
$H_1$ and $H_2$ are as given in (\ref{h1}) and (\ref{h2})
respectively.

\smallskip

\begin{lem}
The Li\' {e}nard equation can be recast as \be \dot{z} =
\{H_1+iH_2, z \}, \ee where $z = \frac{1}{\sqrt{2\alpha}}(\alpha x
- ip)$. The conjugate
 $z^{\ast} = \frac{1}{\sqrt{2\alpha}}(\alpha x + ip)$
and $z$ satisfy
\be
\{z^{\ast}, z \} = -i.
\ee
\end{lem}

{\bf Proof}: Equating real and imaginary part we obtain
$$
\alpha\dot{x} =\alpha \{H_1,x\} + \{H_2,p\}, \qquad
-\dot{p} = \alpha\{H_2,x\} - \{H_1,p\}.
$$
By normalizing these set of we obtain that these equations are
equivalent to $\dot{p}$ and $\dot{x}$ equations. Thus we find our desired
result. Moreover it is easy to check directly that $\{z^{\ast}, z
\} = -i.\quad \Box$

By changing coordinates $(x, p) \to (z, z^{\ast})$ one rewrite the Hamiltonian
equation in complex form in terms of the complex Poisson bracket
\be
\{K,L\} =  -i \big(\frac{\partial K}{\partial z^{\ast}}\frac{\partial L }{\partial z} -
\frac{\partial K}{\partial z}\frac{\partial L}{\partial z^{\ast}}\big).
\ee

\begin{prop}

Suppose the Chiellini integrability condition is satisfied for
the Li\'enard equation of motion. Then Li\'enard equation can be expressed
in complex Hamiltonian form  \be \dot{z}
= \{H_{\Bbb C}, z \} = -i\frac{\partial H_{{\Bbb C}}}{\partial z^{\ast}}, \ee
with  complex coordinates and
complex Hamiltonian function $ H_{\Bbb C} = H_1 + iS$, where
$H_1$ and $S$ are as given in (\ref{h1}) and (\ref{h2})
respectively.

\end{prop}

\section{Concluding remarks}

In this paper we have studied the classic Chiellini integrability condition
$\frac{d}{dx}(\frac{g}{f}) = \alpha(1-\alpha)f(x)$,
and its role
in the construction of Lagrangian and Hamiltonian of the Li\'enard equation. We have formulated the
Chiellini condition using the Jacobi last multiplier, and show that the only equation which satisfies both
the Chiellini condition and Sabatini's isochronicity condition is the Kukles system.
We have re-examined the result by mapping the  Li\'enard equation to harmonic oscillator equation
using the $\alpha$ values of the Chiellini integrability condition.

\smallskip

The Li\'enard equation exhibits many interesting feature of dynamics, we only focus on integrability
or isochronicity aspects. In other words we study the Lagrangians and Hamiltonians of the those Li\'enard system
which have centers. It would be interesting to study Hamiltonian aspects of the Li\'enard equation which
has attractor and other dynamical features, one such work studied dissipative dynamical systems capable of
showing limit cycle oscillations using  canonical perturbation theory \cite{Sagar}.

\section*{Acknowledgement}

The authors wish to thank Professors Tudor Ratiu, Haret Rosu, Andy Hone, Jean Luc Thiffeault and Ogul Esen for
their interest encouragement and valuable suggestions. One of us (PG) wishes to acknowledge
all the members of ICTS, Bangalore, for their gracious hospitality where the last part of this
work was done.

\end{document}